\documentclass{article}
\usepackage{graphicx}
\usepackage{amsmath}

\input{tcilatex}

\begin{document}

\title{Scalar Field Solutions In Colliding Einstein-Maxwell Waves}
\author{M. Halilsoy\thanks{%
mustafa.halilsoy@emu.edu.tr} and I.Sakalli\thanks{%
izzet.sakalli@emu.edu.tr} \\
Physics Department\\
Eastern Mediterranean University, G.Magusa, North Cyprus,\\
Mersin 10 - Turkey.}
\date{\today }
\maketitle

\begin{abstract}
A simple method is presented which enables us to construct scalar field
solutions from any given Einstein-Maxwell solution in colliding plane waves.
As an application we give scalar field extensions of the solution found by
Hogan, Barrab\v{e}s and Bressange.
\end{abstract}

\noindent

Solution generation techniques are the mostly adopted methods in obtaining
new solutions in general relativity. Rarely, however, solutions are found
that hardly can be obtained by simple means. One such example is the
solution obtained by Hogan, Barrab\v{e}s and Bressange (HBB) [1] in the
collision of Einstein-Maxwell (EM) waves. This solution does not belong to
any known family of solutions in this context [2,3]. It has been cast into
the initial value problem within the Ernst formalism [4] and emerges as an
extension of the solution found by Griffiths long ago [5]. The latter
represents collision of an impulsive gravitational wave with an
electromagnetic (em) shock wave. Due to the non-symmetrical initial data
such problems may be labelled as hybrid types. The solution of HBB adds an
impulsive gravitational wave parallel to the incoming em wave and collides
the combination with another impulsive wave. Naturally this solution admits
both the Khan-Penrose [6], as well as the Griffiths limits. We note that the
problem of colliding superposed waves was formulated before [7], but finding
exact solutions remained ever challenging, the HBB solution is therefore
important in this sense.

In this letter by employing a known technique from the stationary axially
symmetric fields [8] in colliding wave, we obtain scalar field extensions of
the HBB solution.

The general metric representing colliding Einstein-Maxwell-Scalar (EMS)
waves is [9],

\begin{equation}
ds^{2}=2e^{-M}dudv-e^{-U}\left[ \left( e^{V}dx^{2}+e^{-V}dy^{2}\right) \cosh
W-2\sinh Wdxdy\right]  \tag{1}
\end{equation}

where the metric functions depend at most on $u$ and $v$. The relevant field
equations are

\begin{equation*}
2U_{\unit{u}u}-U_{u}^{2}+2M_{u}U_{u}=V_{u}^{2}\cosh ^{2}W+W_{u}^{2}+4\varphi
_{u}^{2}+4\Phi _{22}
\end{equation*}

\begin{equation}
2U_{vv}-U_{v}^{2}+2M_{v}U_{v}=V_{v}^{2}\cosh ^{2}W+W_{v}^{2}+4\varphi
_{v}^{2}+4\Phi _{00}  \tag{2}
\end{equation}

\begin{equation*}
2M_{uv}+U_{u}U_{v}=W_{u}W_{v}+V_{u}V_{v}\cosh ^{2}W+4\varphi _{u}\varphi _{v}
\end{equation*}

\begin{equation*}
2\varphi _{uv}=U_{u}\varphi _{v}+U_{v}\varphi _{u}
\end{equation*}

where $\varphi $ is the scalar field and subletters imply partial
derivatives. Now we make a shift (the $M$-shift) in the metric function $M$
according to \ \ \ \ \ \ \ 

\begin{equation}
M\rightarrow M+\Gamma  \tag{3}
\end{equation}

where $\Gamma $ is a function related to the scalar field $\varphi $ through

\begin{equation*}
\Gamma _{u}U_{u}=2\varphi _{u}^{2}
\end{equation*}

\begin{equation}
\Gamma _{v}U_{v}=2\varphi _{v}^{2}  \tag{4}
\end{equation}

\begin{equation*}
\Gamma _{uv}=2\varphi _{u}\varphi _{v}
\end{equation*}

Finding a regular, well-behaved scalar field by this technique is not
guaranteed. As an application we consider the solution by HBB in which the
waves are linearly polarized (i.e. $W=0$). The incoming metrics in the HBB
problem are

\begin{equation}
ds^{2}=2dudv-\left( 1+ku\right) ^{2}dx^{2}-\left( 1-ku\right) ^{2}dy^{2}%
\text{, \ \ \ (Region II)}  \tag{5}
\end{equation}

\begin{eqnarray}
ds^{2} &=&2dudv-\left( \cos \left( bv\right) +\frac{l}{b}\sin \left(
bv\right) \right) ^{2}dx^{2}-\left( \cos \left( bv\right) -\frac{l}{b}\sin
\left( bv\right) \right) ^{2}dy^{2}  \TCItag{6} \\
&&\text{\ \ \ \ \ \ \ \ \ \ \ \ \ \ \ \ \ \ \ \ \ \ \ \ \ \ \ \ \ \ \ \ \ \
\ \ \ \ \ \ \ \ \ \ \ \ \ \ \ \ \ \ \ \ \ \ \ \ \ \ \ \ \ \ \ \ \ \ \ \ \ \
\ \ \ \ \ (Region III)}  \notag
\end{eqnarray}

in which $u$ and $v$ are to be used with the step functions,

\begin{equation*}
u\rightarrow u\Theta \left( u\right) \text{ and }v\rightarrow v\Theta \left(
v\right) \text{ }
\end{equation*}

Here, $k$ and $l$ are the impulsive gravitational wave parameters while $b$
represents the em constant. We note that our coordinate $v$ (in Region III)
is different from the one employed by HBB. (i.e. the relation is $%
v\rightarrow \frac{1}{b}\tan \left( bv\right) $, so that in the limit $%
b\rightarrow 0$ they coincide). The metric functions and the em field
strengths found by HBB are

\begin{equation*}
e^{-U}=F\cos ^{2}\left( bv\right)
\end{equation*}

\begin{equation*}
e^{V}=\frac{1+kuB+\sqrt{1-B^{2}}A}{1-kuB-\sqrt{1-B^{2}}A}
\end{equation*}

\begin{equation}
e^{-M}=\frac{H^{2}}{AB\sqrt{F}}  \tag{7}
\end{equation}

\begin{equation*}
\Phi _{2}=\frac{-k\tan \left( bv\right) B}{AH\sqrt{F}}
\end{equation*}

\begin{equation*}
\Phi _{0}=\frac{b\left[ \left( \frac{l^{2}+b^{2}}{l^{2}}\right) ku\left(
1-B^{2}\right) ^{\frac{3}{2}}+AB^{3}\right] }{BH\sqrt{F}}
\end{equation*}

where

\begin{equation*}
F=A^{2}+B^{2}-k^{2}u^{2}\tan ^{2}\left( bv\right) -1
\end{equation*}

\begin{equation*}
H=AB-ku\sqrt{1-B^{2}}
\end{equation*}

and

\begin{equation*}
A=\sqrt{1-k^{2}u^{2}}\ \ \ \ \ \ \ \ \ B=\sqrt{1-\frac{l^{2}}{b^{2}}\tan
^{2}\left( bv\right) }
\end{equation*}

In order to introduce scalar fields through $M$-shift we observe first that
by introducing new coordinates $\left( \tau ,\sigma \right) $ defined by

\begin{equation*}
\tau =B\cos \left( bv\right) \sqrt{1-A^{2}}+A\sqrt{1-\cos ^{2}\left(
bv\right) B^{2}}
\end{equation*}

\begin{equation}
\sigma =B\cos \left( bv\right) \sqrt{1-A^{2}}-A\sqrt{1-\cos ^{2}\left(
bv\right) B^{2}}  \tag{8}
\end{equation}

the metric function $U$ takes the form

\begin{equation}
e^{-U}=\sqrt{1-\tau ^{2}}\sqrt{1-\sigma ^{2}}  \tag{9}
\end{equation}

This casts the scalar field equation into the form

\begin{equation}
\left[ \left( 1-\tau ^{2}\right) \varphi _{\tau }\right] _{\tau }-\left[
\left( 1-\sigma ^{2}\right) \varphi _{\sigma }\right] _{\sigma }=0  \tag{10}
\end{equation}

which readily admits an infinite class of seperable solutions. We wish to
present two particular solutions.

a) Let

\begin{equation}
\varphi \left( \tau ,\sigma \right) =a_{1}\tau \sigma  \tag{11}
\end{equation}

where $a_{1}$is a constant. The $\Gamma $ function integrates to

\begin{equation}
\Gamma =a_{1}^{2}\left( \tau ^{2}+\sigma ^{2}-\tau ^{2}\sigma ^{2}\right) 
\tag{12}
\end{equation}

This choice of scalar field occurs from both sides of the incoming waves and
it is regular. The em components remain unchanged, same as in the HBB
solution.

b) Let

\begin{eqnarray}
\varphi \left( \tau ,\sigma \right) &=&a_{2}\tanh ^{-1}\left( \frac{\tau
+\sigma }{1+\tau \sigma }\right) ,\text{ \ \ \ \ \ \ \ }\left( u>0,v>0\right)
\TCItag{13} \\
&=&0\text{ , \ \ \ \ \ \ \ \ \ \ \ \ \ \ \ \ \ \ \ \ \ \ \ \ \ \ \ \ \ \ \ \
\ \ \ \ \ }\left( u\leq 0\right)  \notag
\end{eqnarray}

where $a_{2}$ is another constant. The $\Gamma $ function now becomes

\begin{equation}
e^{-\Gamma }=\left( \frac{\left( 1-\tau ^{2}\right) \left( 1-\sigma
^{2}\right) }{\left( \tau +\sigma \right) ^{4}}\right) ^{a_{2}^{2}}  \tag{14}
\end{equation}

In this particular class the scalar field exists only for $u>0,$ which in
the Region II takes the form

\begin{equation}
e^{-\Gamma }=\left( \frac{1-k^{2}u^{2}}{4k^{2}u^{2}}\right) ^{2a_{2}^{2}} 
\tag{15}
\end{equation}

and is well-defined. This solitonic scalar field occurs only in Region II
and IV, while in Region III there is no scalar field. This is another
example of a hybrid type of wave packets whose exact solution is available.
In Region II we have ``gravity + scalar'' wave versus the ``gravity + em''
wave of the Region III.

Finally we remark that by a choice of a scalar field we may set $M=0$. This
amounts to integrating the equations in (4) with $\Gamma =-M$, for a
possible scalar field. This, of course does not guarantee the existence of
such a scalar field, but in some cases it works.

\end{document}